# Tuning of structural phase, magnetic spin order and electrical conductivity in mechanical alloyed material of α-Fe$_2$O$_3$ and α-Cr$_2$O$_3$ oxides

R.N. Bhowmik[*a], K. Venkata Siva[a], V.R. Reddy[b], and A.K. Sinha[c,d]

[a]Department of Physics, Pondicherry University, R. Venkataraman Nagar, Kalapet, Pondicherry-605014, India

[b]UGC-DAE Consortium for Scientific Research, Khandwa Road, Indore - 452 001, India

[c]HXAL, SUS, Raja Ramanna Centre for Advanced Technology, Indore- 452013, India

[d]Homi Bhabha National Institute, Anushakti Nagar, Mumbai -400 094 India

[*]Corresponding author: Tel.: +91-9944064547; Fax: +91-413-2655734

E-mail: rnbhowmik.phy@pondiuni.edu.in

**Abstract:** α-Fe$_2$O$_3$ and α-Cr$_2$O$_3$ has been mechanical alloyed to prepare Fe$_{1-x}$Cr$_x$O$_3$ oxides for x = 0.2-0.8. Synchrotron X-ray diffraction and Raman spectra have shown inhomogeneous structure of α-Fe$_2$O$_3$ and α-Cr$_2$O$_3$ phases in as-alloyed samples. The as-alloyed samples have shown soft ferromagnetic properties with signature of two Morin transitions. The heat treatment of as-alloyed samples has homogenized structure and successfully incorporated the Cr atoms into the lattice sites of Fe atoms in α-Fe$_2$O$_3$. The magnetic and electrical properties have been modified in the heat treated samples. For example, canted antiferromagnetic order has been appeared as an effect of heat treatment, irrespective of the Cr content in Fe$_{1-x}$Cr$_x$O$_3$. The magnetic field induced spin flop transition has been observed at a critical magnetic field that depends on Cr content in the system. The Mössbauer spectrum at room temperature has been fitted with two sextets. The variation of Mössbauer parameters suggest a distribution of magnetic spin order between Fe and Cr ions in the rhombohedral structure of Fe$_{1-x}$Cr$_x$O$_3$. The electrical conductivity, derived from current-voltage characteristics of the heat treated samples, has been enhanced by increasing Cr content in α-Fe$_2$O$_3$ structure. The experimental results have been explained based on the theoretical models available in literature.

# 1. Introduction

The solid solution of α-$Fe_2O_3$ and α-$Cr_2O_3$ oxides has drawn a lot of theoretical and experimental interest for the development of new magnetic semiconductor oxides with novel magnetic, electronic, optical, and magneto-electronic properties [1-5]. The α-$Cr_2O_3$ (chromia) and α-$Fe_2O_3$ (hematite) are isomorphs of corundum structured (space group $R\bar{3}c$) α-$Al_2O_3$ (alumina). The metal ions are located at either slightly above or below of the rhombohedral planes. The metal ions ($Fe^{3+}/Cr^{3+}$) randomly occupy $Al^{3+}$ sites in corundum (rhombohedral) structure and alternating layers of metal ions are separated by the layers of oxygen ($O^{2-}$) ions along c (001)-axis [6-8]. Despite identical lattice structure, magnetic structure in α-$Fe_2O_3$ and $Cr_2O_3$ is different. α-$Fe_2O_3$ is a typical canted antiferromagnet (CAFM) in the temperature range 960 K (Neel temperature $T_N$) to Morin transition (a first-order magnetic transition) at $T_M \sim 260$ K, below which bulk α-$Fe_2O_3$ behaves as uniaxial antiferromagnet AFM [1]. In the CAFM state, spin moment of $Fe^{3+}$ ions form in-plane ferromagnetic (FM) order and inter-planes antiferromagnetic (AFM) order with small canting due to anisotropic Dzyaloshinskii-Moriya interactions. Below $T_M$, the spins in bulk α-$Fe_2O_3$ are oriented along (001) axis (out of plane). The magnetic properties of bulk $Cr_2O_3$ are significantly different, where magnetic spin order is highly magnetic field sensitive below its $T_N$ at 308 K [9]. In the absence of external magnetic field, $Cr^{3+}$ spins form AFM order along (001) direction with a sequence of up (+) and down (-) order + − + −. In the presence of magnetic field, the spins of $Cr^{3+}$ ions flop along in-plane direction at a typical field value of 6 Tesla, which is slightly less than the spin-flop field of 6.75 Tesla for bulk α-$Fe_2O_3$ [9]. The Morin transition is absent in α-$Cr_2O_3$. The alloying of α-$Fe_2O_3$ and α-$Cr_2O_3$ oxides, thus, exhibited unusual magnetic properties [4-5, 10-11]. However, application of α-$Fe_2O_3$ in electronic devices is limited due to low charge carrier density, low carrier mobility, and high electron–hole recombination rate [12]. Various theoretical models have been proposed to enhance electronic charge transport in α-Fe$_2$O$_3$ system by metal doping [6-7] and these theoretical models need experimental evidences.

The experimental work on modified magnetic spin order and electronic conductivity in the alloy of α-Fe$_2$O$_3$ and α-Cr$_2$O$_3$ oxides has not been reported much. A detailed study of the structure, magnetic, and electronic properties of α-Fe$_{2-x}$Cr$_x$O$_3$ system for a wide range of Cr content at different stages of their preparation using mechanical alloying and heat treatment has been presented in this work. Attempt has been made to establish a correlation between structure and properties by using the results of Synchrotron X-ray diffraction, magnetization, Mössbauer spectroscopy and electrical conductivity measurements.

2. **Experimental**

**2.1 Sample preparation**

The Cr doped hematite system Fe$_{1-x}$Cr$_x$O$_3$ (x = 0.2-0.8) was prepared by mechanical alloying of the stoichiometric amounts of α-Fe$_2$O$_3$ and α-Cr$_2$O$_3$ powders for 80 h in air using FRITISCH (Pulverisette 6, Germany) planetary mono miller. The stainless steel balls (∅ 10 mm) were used for alloying process. The material to ball mass ratio was maintained at 1:5. The mechanical alloying was intermediately stopped for mixing and better homogenization of the product. The alloyed material was pressed into pellets (ASP), which were heated at 800 °C (for one composition) and 1000 °C (for all compositions) in air for 12 h. The as-alloyed samples of Fe$_{1-x}$Cr$_x$O$_3$ were denoted as FeCr2MA, FeCr4MA, FeCr5MA, FeCr6MA, and FeCr2MA for x = 0.2, 0.4, 0.5, 0.6, and 0.8, respectively. The heat treated samples at 1000 °C (A10 samples) were denoted as FeCr2MA10, FeCr4MA10, FeCr5MA10, FeCr6MA10, and FeCr8MA10 for x = 0.2, 0.4, 0.5, 0.6, and 0.8, respectively.

**2.2 Sample characterization and measurements**

Lattice structure of the samples has been studied by recording high quality synchrotron X-ray diffraction (SXRD) patterns at room temperature. The SXRD patterns were recorded at angle dispersive x-ray diffraction (ADXRD) beam line (BL-12) at Indus-2 synchrotron source, RRCAT, India using an image plate (MAR 345) area detector. The wave length ($\lambda$ = 0.79923 Å) and the sample to detector distance were accurately calibrated by performing XRD on $LaB_6$ NIST standard using same set up. Raman spectra of the samples were recorded in the wave number range 100–1000 $cm^{-1}$ using microscope (Renishaw, UK). The dc magnetization in the temperature range 10-350 K was measured using physical properties measurement system (PPMS-EC2, Quantum Design, USA). The temperature dependence of magnetization (M (T)) curves were measured following conventional zero field cooled (ZFC) and field cooled (FC) modes. In ZFC mode, the sample was cooled in the absence of external magnetic field from room temperature (~ 300 K) to 10 K and MZFC(T) curve was recorded in the presence of an applied magnetic field during increase of the sample temperature up to 350 K. After reaching the temperature at 350 K, the sample was field cooled (using same field as used in MZFC(T) measurement) down to 10 K and MFC(T) curve was recorded during increase of the sample temperature up to 350 K. The magnetic field dependent magnetization (M(H)) curves were recorded within field range of 0 to ± 70 kOe. $^{57}$Fe Mössbauer measurements were carried out in transmission mode with a $^{57}$Co (Rh matrix) radio-active source in constant acceleration mode using a standard PC-based spectrometer. The spectrometer was calibrated with natural Fe foil at room temperature. The Mössbauer spectra were analysed with NORMOS-DIST/SITE program for the estimation of spectral parameters. The High Resistance Meter (Keithley 6517B) was used to study the current vs. voltage (I-V) characteristics of the samples at room temperature. This meter provides a precise voltage and current sourcing and measurements for high resistive materials. The I-V characteristics were measured by sandwiching the disc shaped (⌀ 10-12 mm, thickness 0.5-1.2 mm) samples between Pt electrodes using a home-made sample holder to make a Pt/M/Pt structure (M: material). The bias voltage was swept within ±200 V with step size 1 V.

## 3. Results and discussion

### 3.1 Structural properties

Fig.1 shows SXRD patterns of the as-alloyed (without post heat treatment) samples. The patterns are consistent to rhombohedral (corundum) structure and splitting of the peaks shows incomplete alloying of the α-$Fe_2O_3$ and α-$Cr_2O_3$ phases. The peak component at higher 2θ corresponds at α-$Cr_2O_3$ phase (smaller lattice constant) and peak component at lower 2θ corresponds to α-$Fe_2O_3$ phase (higher lattice constant) [4]. The decrease of peak intensity for α-$Fe_2O_3$ phase is accompanying with an increasing peak intensity of α-$Cr_2O_3$ phase when Cr content increases in the composition. The SXRD patterns (Fig. 2) of A10 samples (heated at 1000 °C) show single phased corundum structure. Structural refinement was performed by assigning metal atoms and oxygen atoms in 12c and 18e Wyckoff positions, respectively. The SXRD patterns was fitted by adopting Rhombohedral structure with space group $D_{3d}^6$ ($R\bar{3}c$), where metal (Fe/Cr) and O atom occupy the positions (0, 0, $z_M$) and ($x_O$, 0, $\frac{1}{4}$), respectively [13]. The structural refinement started with initial parameters values $Z_M = 0.35$ and $X_O = 0.31$. Table 1 and Table 2 show the fit parameters (atomic positions, cell parameters, crystallite size and phase fraction) for the as-alloyed and (1000 $^0$C) heat treated samples, respectively. The positional parameters have shown a considerable change on increasing the Cr content (x) in the $Fe_{1-x}Cr_xO_3$ system. The $Z_{Fe}$ (for Fe atoms) and $Z_{Cr}$ (for Cr atoms) values are found smaller than the value 0.35498 used for pure α-$Fe_2O_3$ [13]. For as-alloyed samples, $Z_{Cr}$ values are relatively smaller (due to smaller cation size) in comparison to the values of $Z_{Fe}$. The $Z_{Fe}$ values are nearly unchanged for the Cr content x = 0.2 to 0.5 and rapidly decreased for x ≥ 0.6. The $Z_{Cr}$ showed a gradual decrease with the increase of x. The position parameter of Oxygen atom ($X_O$) is relatively larger in the α-$Fe_2O_3$ phase and decreased linearly (within error bar), except a noticeable decrease for x = 0.6. The $X_O$ values in α-$Cr_2O_3$ phase are relatively smaller, but it increased with the increase of Cr content in the system. In the heat treated samples, $Z_{Fe}$ values linearly decreased on increasing the Cr content and $X_O$ values also decreased with a noticeably small value for x = 0.5. The lattice parameter (*a*) of the as-alloyed materials decreased in both the phases (*a*: 5.0384 Å to 5.0171 Å for α-Fe$_2$O$_3$ phase and 4.9641 Å to 4.9456 Å for α-Cr$_2$O$_3$ phase). The cell volume in both the phases decreased rapidly for x up to 0.5 and thereafter, changes are minor. The fraction of α-Fe$_2$O$_3$ phase decreases by increasing the α-Cr$_2$O$_3$ phase in the as-alloyed material, except an intermediate increase of α-Fe$_2$O$_3$ phase at x = 0.6 when x value increases from 0.2 to 0.8. In case of the (1000 $^0$C) heat treated samples, the structural parameters are stabilized with smaller $\chi^2$ values (1.3-2.8). The $Z_{Fe}$ values and cell parameters (including c/a ratio) systematically decreased on increasing the Cr content in the system. This is consistent to the substitution of Fe$^{3+}$ ions with larger ionic radius (0.645 Å) by Cr$^{3+}$ ions with smaller ionic radius (0.615 Å) [6, 14-15]. The *c/a* ratio lie in the range of Ga doped α-Fe$_2$O$_3$ system [16]. The $X_O$ also decreased, except a noticeable reduction for the composition x = 0.5. This is related to structural disorder due to composition variation [17-18]. Crystalline size (~ 30-35 nm) of the heat treated samples was calculated using Debye-Scherrer formula). It is not much composition dependent for the samples with identical heat treatment.

Fig.3 (a-f) represents Raman spectra of the as-alloyed and heat treated (A10) samples. The extra phase (∗ marked) in as-alloyed samples appeared due to non-reacted Cr$_2$O$_3$ phase [19]. The absence of any extra phase for A10 samples confirms a complete solid solution of Cr$_2$O$_3$ and α-Fe$_2$O$_3$ phases [20-21]. Raman spectra of A10 samples with six phonon modes (one A$_{1g}$ mode, four E$_g$ modes and one Raman inactive E$_u$ mode) suggest rhombohedral structured Cr doped α-Fe$_2$O$_3$ system [5]. Table 3 shows the band positions in A10 samples. An additional peak at 665-670 cm$^{-1}$ in the spectra of Fe$_{1-x}$Cr$_x$O$_3$ samples with respect to the spectrum of α-Fe$_2$O$_3$ [20] and α-Cr$_2$O$_3$ [22] suggest a modified spin-lattice structure. The band (peak) position (within simple harmonic approximation) obeys the equation $\bar{v} = \frac{1}{2\pi c}\sqrt{\frac{k_{M-O}}{\mu_{M-O}}}$.

The peak position ($\bar{\nu}$) in Raman spectrum depends the spring constant ($k_{M-O}$) of metal-oxygen bond (M-O) with length $r_{M-O}$ and $\mu_{M-O}$ is effective mass. The spring constant ($k_{M-O}$), being second order derivative of the coulomb potential ($V(r_{MO})$), is proportion to $\frac{1}{r_{MO}^3}$. The bond length ($r_{M-O}$) is expected to be less by doping of $Cr^{3+}$ ions and supported from the decrease of cell parameters. Hence, spring constant ($k_{M-O}$) is expected to increase with the increase of Cr content and it is to reflect in band shift. However, Table 3 shows shift of the band positions to lower values, except the $E_g(5)$ band whose position increases on increasing Cr content. The bands carry information about its internal structural changes. Especially the band position at lower frequencies ($E_g(1)$, $E_g(3)$, $E_g(4)$) represent symmetric stretching of O atoms relative to metal (M) atoms in the in-planes and the $A_{1g}$ band correlates the movement of cations along out of plane direction of the Rhombohedral structure [5, 16]. The IR-active Eu(LO) band is theoretically forbidden. It reflects local lattice disorder and internal strain in the system. The decrease of Eu(LO) band position indicates an adjustment of Fe-O and Cr-O bonds along in-plane and out of plane directions to reduce lattice disorder in the Cr doped α-$Fe_2O_3$ system.

**3.2 dc magnetization**

Fig. 4 (a-e) shows the temperature dependence of low field (100 Oe) magnetization curves for the as-alloyed samples. A divergence between MZFC(T) (black box symbol) and MFC(T) (red circle symbol) curves is noted for all samples below the highest measurement temperature 350 K. The features of M(T) curves in the as-alloyed samples of $Fe_{2-x}Cr_xO_3$ are drastically different from that in mechanical milled α-$Fe_2O_3$ [23] and α-$Cr_2O_3$ [24], and chemical routed Cr doped α-$Fe_2O_3$ samples [5]. The M(T) curves of the as-alloyed samples do not show a strong signature of Morin transition as observed at 260 K in bulk α-$Fe_2O_3$ [23]. However, prominent slope change in M(T) curves at about 60 K and 300 K (marked by vertical lines) indicated weak features of the Morin transitions in as-alloyed samples. It is comparable to the features of Ga doped α-$Fe_2O_3$ system [20]. The temperature dependence of MZFC curves (Fig. 4(f)), normalized (at 300 K) MFC curves (Fig. 4(g)), difference (ΔM) between MFC and MZFC curves (Fig. 4(h)), and temperature derivative of the ΔM curves (Fig. 4(i)) are plotted to clarify the changes in M(T) curves. The normalized MFC curves showed two local minima correspond to the slope changes in MZFC curves and an intermediate broad maximum that ended with a distinct kink in the MFC curves at about 100 K. This shows formation of some meta-stable magnetic states during field cooling process. Although ΔM (T) curves do not provide good information for existence of Morin transitions, but temperature derivative of the ΔM curves showed shoulders near to the Morin transitions that marked in MZFC (T) curves.

The low field (100 Oe) MZFC(T) and MFC(T) curves (Fig.5(a-e)) of the heat (1000 $^0$C) treated samples showed the features, which are different from as-alloyed samples. The signature of Morin transition is seen at ~ 247 K in the MZFC(T) curve of FeCr2MA10 sample and a weak signature is noted at about 245 K for FeCr4MA10 sample. MFC(T) curve of the FeCr2MA10 sample showed a high magnetic state below the Morin transition and the character is different from the features of high Cr doped samples. The separation between MZFC and MFC curves, which extended down to 10 K, also decreases on increasing the Cr content. The magnetization up turn below 100 K represents the effect of frustrated surface spins in AFM nanoparticles [25]. The MZFC(T) curves at different magnetic fields were measured for the samples FeCr2MA10 (Fig. 5(f)) and FeCr5MA10 (Fig. 5(g)). The M(T) curves were normalized by its value at 350 K. There is no signature of Morin transition at the higher magnetic fields for the FeCr2MA10 sample. In fact, Morin transition in α-Fe$_2$O$_3$ arises due to a competition between magnetic dipolar anisotropy ($K_{MD}$) among magnetic Fe$^{3+}$ ions (which controls in-plane spin order above $T_M$) and the single ion anisotropy ($K_{SI}$) from higher orders spin-orbit coupling of Fe$^{3+}$ ion (which controls the out of plane spin orientation at temperatures below $T_M$) [1, 20]. The anisotropy in α-Cr$_2$O$_3$ is minor due to cancellation of the opposite signed and nearly equal strength of dipolar contribution ($K_D$) and crystal field contribution ($K_x$). Hence, suppression of $T_M$ is expected in single phased $Fe_{2-x}Cr_xO_3$ system at higher Cr content. On the other hand, a kink (magnetic field independent) appeared at ~ 51 K and 55 K for FeCr2MA10 and FeCr5MA10 samples, respectively. Such low temperature magnetic anomaly is attributed to freezing of a fraction of surface spins in AFM nanoparticles [9, 22]. The normalized MZFC(T) curves for FeCr2MA10 sample showed a minor increment by increasing the field up to 2 kOe. This indicates a strong AFM spin order [23]. On the other hand, MZFC(T) curves of FeCr5MA10 sample are appreciably changed at higher fields and it can be attributed to the field sensitive flipping of $Cr^{3+}$ spins [9].

The M(H) curves are measured to understand the field induced spin order/spin flop transition in the samples. M(H) curves of the as-alloyed samples (Fig.6 (a-e)) show soft FM character at all measurement temperatures (10 K-350 K). The existence of small loop and magnetic irreversibility with lack of magnetic saturation for field up to 70 kOe is clarified in the insets of Fig. 6(a-e). The M(H) curves at different temperatures appeared to overlap in the full scale of magnetic field, but the difference is shown in the inset of Fig. 6(c) between M(H) curves at 10 K and 300 K. The It is noted that the magnetic field below which M(H) loop at 10 K showed irreversibility increases for increasing the Cr content x from 0.2 to 0.5 and again decreases on further increase of x up to 0.8. The M(H) loop at 10 K for x = 0.5 is not closed for field up to 70 kOe. The exchange bias effect in as-alloyed samples is studied by measurement of FC-M(H) loops (Fig. 6 (f-h)) at 10 K after field cooling the samples from 300 K under 70 kOe. The FC-M(H) loop of the samples showed a shift along negative field direction with respect to ZFC loop. The exchange bias field ($H_{exb} = H_0^{FC} - H_0^{ZFC}$), i.e., shift of the center of FC ($H_0^{FC} = (H_{C1}^{FC} + H_{C2}^{FC})/2$) loop with respect to the center of ZFC loop ($H_0^{ZFC} = (H_{C1}^{ZFC} + H_{C2}^{ZFC})/2$)), was found to be - 58 Oe for FeCr2MA, - 78 Oe for FeCr4MA, and -94 Oe for FeCr8MA samples. At the same time, remanent magnetization showed a shift ($M_{exb} = M_0^{FC} - M_0^{ZFC}$) along positive direction by the amount + 0.282 emu/g, + 0.154 emu/g, and + 0.0869 emu/g for FeCr2MA, FeCr4MA, and FeCr8MA samples, respectively. A comparative M(H) curves (Fig. 7 (a-c)) at temperatures 10 K, 200 K and 300 K for the composition $Fe_{1.5}Cr_{0.5}O_3$ has shown that magnetic moment and FM order of as-alloyed (ASP) sample are reduced in heat treated samples. The M(H) loop is widened in the heat treated samples in comparison to a rod shaped narrow loop in the as-alloyed sample. Magnetic features of the heat treated samples are identical, although A8h12 sample (heated at 800 $^0$C) showed slightly higher moment and loop area in comparison to the A10h12 sample (heated at 1000 $^0$C). This is clarified in Fig. 7(c). FC-M(H) loop of the A10h12 sample at 10 K also showed exchange bias shift with respect to the ZFC loop (Fig. 7(d)). The shift values are $H_{exb}$ ~ -128.6 Oe and $M_{exb}$ ~ + 0.00257 emu/g). The exchange bias effect in Fe-Chromia phase was modelled [29] as the magnetic exchange coupling between AFM core of α-$Cr_2O_3$ and ferrimagnetic shell of α-$Fe_{0.40}Cr_{1.60}O_{2.92}$. The observation of exchange bias effect in our as-alloyed and heat treated samples, despite of having different magnetic features, can be attributed to magnetic coupling at the interfaces with different spin order and strength of AFM and FM interactions. It can be of core-shell structure or non-uniform distribution of Cr and Fe moments in the Rhombohedral planes [5, 25-28]. In Fig. 8, we have compared the temperature variation of magnetic field parameters at different stages of material preparation for the composition $Fe_{1.5}Cr_{0.5}O_3$. The unusual feature is the shift of ZFC loop ($H_{exb}^{ZFC} \neq 0$, $M_{exb}^{ZFC} \neq 0$) for all the samples (without any pre-field cooling process) with reference to the origin of M-H plane at all measurement temperatures. Such shift in the ZFC-M(H) loop could be attributed to magnetic modulation of the FM and AFM exchange interactions at the interfaces of Cr doped hematite system [29] and noted in Ti doped α-$Fe_2O_3$ [30]. It is noted that the $H_{exb}^{ZFC}$ in ASP sample is less temperature dependent (10K-350 K) and the values varied within + 5 Oe to +15 Oe. On the other hand, $H_{exb}^{ZFC}$ values are relatively large and negative at low temperatures ($H_{exb}^{ZFC}$ ~ - 330 Oe for A8h12 sample and - 837 Oe for A10h12 sample at 10 K) that approaches to the zero value (for A10h12 sample) or crosses to positive value (for A8h12 sample) on increasing the temperature to 300 K and above. Interestingly, $H_C$ of the ASP sample was found in the 270-500 Oe and it decreased at higher temperatures (a typical signature of temperature activated decrease of FM order) in contrast to an increase of coercivity with temperature in heat treated samples (a typical feature of increasing canted FM order). The coercivity ($H_C$) is much enhanced in the material by heat treatment and it is the largest (3380-3920 Oe) for A8h12 sample. $M_R$ of the samples showed temperature dependent behaviour, similar to that for $H_C$. The exchange bias shift of remanent magnetization ($M_{exb}^{ZFC}$) is positive (i.e., shift of the ZFC loop is along positive magnetization axis) for all these three samples. The magnitude of $M_{exb}^{ZFC}$ is small and nearly temperature independent for ASP sample, where as $M_{exb}^{ZFC}$ is appreciably large for the heat treated samples and it is decreased on increasing the temperature from 10 K to 350 K. The greater exchange bias shift in heat treated sample than the as-alloyed sample suggests increasing AFM interactions that enhanced cooling field induced interfacial coupling between FM and AFM layers [28].

The increase of AFM interactions with the increase of Cr contentment in heat treated (1000 $^0$C) samples can be realized from decreasing area of the M(H) loop at 300 K (Fig. 9 (a-e)). M(H) loop of the FeCr2MA10 sample shows features of a typical CAFM, whereas the features of M(H) loop in FeCr8MA10 sample are comparable to that in AFM nanoparticles of $Cr_2O_3$ [9]. A systematic change of magnetic spin order in A10 samples with increasing Cr content has been understood by analysing the initial M(H) curves (H: 0-70 kOe). The M(H) curves at 10 K (Fig. 9(f-j)) have shown a non-conventional increment of magnetization (up curvature at lower fields and a linear/down curvature increase at higher fields). The difference in magnetic spin order from low field regime to high field regime (spin flop transition) is reflected in the field derivative of magnetization (dM/dH) curves (blue colour curves in Fig. 9(f-j)). The peak position of dM/dH vs. H curves represents spin flop field ($H_{sf}$) and it confirms a changing spin order from second order to first order in corundum structured AFM oxides [31-32]. The Arrott plot $\frac{H}{M} = \beta M^2 + \alpha$ of the M(H) curves (presented by $M^2$ vs. H/M curves in Fig.9 (k-o)) is a powerful technique to distinguish first-order phase transition from second order phase transition below Curie point ($T_C$) of a FM [33-34]. According to Banerjee criterion [34-35], $\beta$ is positive for second-order and negative for first order phase transition. An extrapolation of the linear or polynomial fit of the high field $M^2$ vs. H/M curve on the positive value of $M^2$ axis at H/M = 0 corresponds to $M_S^2(T)$. The $M^2$ vs. H/M plot showed an upward curvature with positive slope ($\beta$) at lower fields for FeCr2MA10 sample and the slope transforms into negative value for the samples with higher Cr content. In the present material, first order magnetic transition is possible due to a local distortion in lattice structure by coupling with spin order. The Raman active band at about 665-670 cm$^{-1}$ suggests a strong spin-lattice coupling in the CAFM phase of metal doped α-$Fe_2O_3$ systems [5, 16, 20]. A change of slope values in the Arrot plot implies a change in the ratio of AFM and FM interactions, where FM order dominates for the samples at low Cr content and AFM order of Cr spins dominates for the samples at high Cr content.

Now, we summarize the magnetic parameters calculated from M(H) curves of the as-alloyed samples (Fig. 10(a-b)) and heat treated (A10) samples (Fig. 10(c-f)). The variation of coercivity ($H_C$) for as-alloyed samples is a bit scattered due to heterogeneity in structural phase of the samples. The scattering is observed mostly for the intermediate compositions x = 0.4-0.6. Roughly, coercivity ($H_C$) at selected temperatures showed an increasing trend on increasing Cr content in the as-alloyed samples. The remanent magnetization ($M_R$) decreased (Fig. 10(b)) with the increase of Cr content in the as-alloyed samples, despite an intermediate fluctuation for the composition at x = 0.5. Both $H_C$ and $M_R$ have shown a general decreasing trend with the increase of measurement temperature from 10 K to 300 K. This behaviour is in contrast to the increase of $H_C$ (Fig. 10(c)) and $M_R$ (Fig. 10(d)) in A10 samples on increasing the measurement temperature (10 K -300 K). However, $H_C$ and $M_R$ both decreased with the increase of Cr content in A10 samples. This indicates an increasing AFM spin order in the Cr doped hematite system. At the same time, FM component increases in the CAFM state as the measurement temperature increases from 10 K to 300 K and above. It can be understood from the variation of spontaneous magnetization ($M_S$) in FeCr2MA10 and FeCr4MA10 samples (Fig. 10(e)), where a competitive magnetic spin order between the low temperature surface spin order and high temperature CAFM spin order is observed in the temperature range 100 K-150 K. The spin flop field ($H_{sf}$), an important parameter for changing magnetic spin order from low magnetic state (AFM) to high magnetic state (CAFM), varied on the variation of both Cr content and measurement temperature of the A10 samples (Fig. 10(f)). The $H_{sf}$ at 10 K shifts to higher values for the increase of Cr content in the material and the values are comparable to that reported for $Cr_2O_3$ system [36]. The $H_{sf}$ values are temperature dependent and showed a general decreasing trend with increase of measurement temperature. A non-monotonic increase of the $H_{sf}$ was noted for the samples (FeCr2MA10 and FeCr5MA10) with intermediate composition, which exhibited higher values in the temperature range 150-200 K.

### 3.3 Mössbauer spectra

Fig. 11 presents Mössbauer spectra of the as-alloyed (ASP) and heat treated at 1000 $^0$C (A10) samples. All the samples showed six-finger spectrum and fitted by Lorentzian shaped pattern with line intensity ratio 3:2:1:1:2:3. The spectral line parameters (Line width (Γ), Isomer shift (IS), Quadrupole splitting (QS), Hyperfine field ($B_{hf}$), and Area (%) of sextets) for different samples are shown in Table 4. The spectrum of FeCr2MA10 sample (low Cr content) is fitted with a single sextet with $B_{hf}$ ~ 48.25 Tesla, QS ~ -0.20 mm/s and IS~ 0.36 mm/s. These values are closely matched with hematite structure [8, 20]. The spectra for ASP and rest of the A10 samples matched with two sextet components. The Mössbauer spectra fitted with two sextets with different $B_{hf}$ shows a non-uniform distribution of Cr and Fe spins in the lattice structure. The first sextet with higher values of $B_{hf}$ (~50-51 Tesla) is attributed to Fe sites with less number of low magnetic Cr ions as nearest neighbours and the second sextet with smaller $B_{hf}$ (~ 47-49 Tesla) corresponds to Fe sites with more number of Cr ions as nearest neighbours [5, 8, 10]. Such distribution of Fe and Cr ions in Cr doped α-$Fe_2O_3$ system suggests the formation of Fe rich and Cr rich environments in rhombohedral structure [20, 37]. The structural heterogeneity may also arise due to core-shell type spin structure in nano-crystallites particles [25, 29]. The $B_{hf}$ values of both the components are nearly unaffected by the variation of Cr content in as-alloyed samples and also for A10 samples. The variation of hyperfine field (magnetic interactions) at Fe sites on increasing the Cr content is not consistent to a monotonic decrease of the dc magnetization in the system. The Mössbauer spectra provide evidence of long ranged spin order at Fe sites (microscopic level) and it does not differentiate much after metal doping in hematite structure [20]. On the other hand, dc magnetization provides an average bulk (macroscopic) value of the magnetic and non-magnetic environments in the material. In case of ASP samples, relative area of the first sextet decreases by increasing the area of second sextet with the increase of Cr content. In case of A10 sample, relative area of the first (second) sextet initially decreases (increases) on increasing the Cr content up to x = 0.5 and then increases (decreases) on further increase of the Cr content within the limit of reasonably good $\chi^2$ values. The line width values in our Cr-Fe oxide samples are found wider than that in the spectrum of bulk α-$Fe_2O_3$ [37]. The line width does not vary monotonically with the increase of Cr content in the system. This shows introduction of magnetic disorder α-$Fe_2O_3$ structure after doping of Cr atoms. Another important parameters, IS and QS, are useful to understand the changes of electronic charge environment in the material. The isomer shift (IS) arises due to variation of the charge density of s-electrons at the environment of probed Nucleus with respect to source of γ-radiation. The Quadrupole splitting (QS) represents the interaction between energy levels of the probed nucleus and surrounding electric field gradient produced by a non-spherical (asymmetric) electronic charge distribution in the nucleus. The negative value of QS indicates the flattening of nucleus along the spin axis. The IS and QS values are found in the range 0.36-0.38 and -0.18 to -0.25 mm/s, respectively. There are some changes in the values of IS and QS with the variation of Cr content in the samples. Especially, IS values correspond to second sextet of the A10 samples systematically decreased from 0.40 to 0.36 mm/s with the increase of Cr content from 0.4 to 0.8. The IS and QS values in the present samples are within the reported values for nanostructured α-$Fe_2O_3$ and Cr doped α-$Fe_2O_3$ systems with average +3 charge state of $Fe^{3+}$ ions [1]. Subsequently, charge state of the Cr ions is assumed to +3 charge state to maintain total charge neutrality in the samples [36]. The IS and QS values for $Fe^{2+}$ charge state are relatively larger [38]. This is due to the fact that shielding effect of d electrons on the s-electrons at the environment of Nucleus is less for the $Fe^{3+}$ ions ($3d^5$) in comparison to that for the $Fe^{2+}$ ions ($3d^6$). However, there is a possibility of fractional ($0 \leq \delta \leq 1$) electronic charge transfer between Fe ($Fe^{(3-\delta)+} \leftrightarrow Fe^{(3+\delta)+}$) and Cr ($Cr^{(3+\delta)+} \leftrightarrow Cr^{(3-\delta)+}$) ions through $Fe^{3+}$-$O^{2-}$-$Fe^{3+}$, $Cr^{3+}$-$O^{2-}$-$Cr^{3+}$ and $Fe^{3+}$-$O^{2-}$-$Cr^{3+}$ superexchange paths in the Cr doped α-$Fe_2O_3$ structure. It will be understood using electrical conductivity of the samples.

**3.4 Electrical conductivity**

Electrical conductivity of the as-alloyed samples is not measured in order to avoid additional heterogeneity effect, especially the effect of bi-phased structure. Electrical field (E) dependence of current density (J) of the 1000 $^0$C heat treated (A10) samples have been recorded by sweeping bias voltage within ± 200 V. The J-E curves (Fig. 12 (a)) showed non-linear behaviour and break down voltage of the samples appears to be high. Such materials could be useful in high power devices [39]. The non-linear I-V feature is clarified in the inset of Fig. 12(a) with current density (J) in log scale. The current density (J) at a specific electric field increases on increasing the Cr content in $Fe_2O_3$ system. The electrical conductivity ($\sigma$) has been calculated by applying the formula $J = \sigma E$ for electric field at 50 kV/m, 100 kV/m and 150 kV/m. Electrical conductivity in Fig. 12(b) has shown a rapid increase by increasing the Cr content. Many factors during alloying of the metal oxides, e.g., composition, structural heterogeneity and intrinsic defects, can control the electrical properties. Especially, the role of defects has been given highlighted for tuning magnetic and electrical properties in $\alpha$-$Fe_2O_3$ [40] and $\alpha$-$Cr_2O_3$ [41-42] and $\alpha$-$Fe_{2-x}Cr_xO_3$ [43]. Based on electron transfer theory, Iordanova et al. [40-41] modelled electronic charge transport in $\alpha$-$M_2O_3$ (M = Fe, Cr) as the alternation of valence state of the metal (M) ions via oxygen bridges (superexchange paths, as mentioned earlier), viz., $M^{3+} + e \rightarrow M^{2+}$ (electron transfer) or $M^{3+} + h \rightarrow M^{4+}$ (hole transfer). The calculations showed that hole mobility in $\alpha$-$Cr_2O_3$ is more than three orders of magnitude larger than the electron mobility in and out of (001) plane directions. In contrast, three orders of magnitude anisotropy in both electron and hole mobility was found between in and out of (001) plane directions of $\alpha$-$Fe_2O_3$. The low conductivity along (001) direction was attributed to hole transfer by $Fe^{3+} + h \rightarrow Fe^{4+}$ process. The difference of charge mobility in and out of (001) plane direction arises from internal reorganization energy for electron-transport relative to hole-transport processes. The electronic coupling between the Cr-3d and $O^{2-}$ states for charge-transfer processes was found smaller in $\alpha$-$Cr_2O_3$ than that between the Fe-3d and $O^{2-}$ states in $\alpha$-$Fe_2O_3$. In the presence of minor defects in the lattice structure, the electrons or holes, while hopping between two sites, can be trapped at the defect sites or M sites. This brings distortion to its surroundings and forms small polarons in $\alpha$-$M_2O_3$ structure. Lebreau et al. [42] studied electronic and magnetic properties for non-defective and defective structures of $\alpha$-$Cr_2O_3$. Different point defects, namely, Cr vacancy, Cr Frenkel defect (composed of an interstitial Cr atom and a Cr vacancy), and O vacancy were assumed. It has been suggested that Cr vacancy and Cr Frenkel defect can induce extra localized level inside the band gap and it results in enhancement of electrical conductivity. The calculations by Iordanova et al. [40-41] showed that the reorganization energy of the electrons or holes is independent of the electron-spin coupling in α-$M_2O_3$ structure. On the other hand, the calculations by Lebreau et al. [42] suggested that magnetic moments on surrounding atoms, especially the first nearest neighbours, are affected by the defects in α-$Cr_2O_3$. α-$Fe_2O_3$ is an anisotropic insulator with higher magnetic moment (5$\mu_B$ per $Fe^{3+}$ ion) and α-$Cr_2O_3$ is a symmetric insulator with low moment (3$\mu_B$ per $Cr^{3+}$ ion) [7]. The DFT calculation by Zs. Rak and D. W.Brenner [44] suggested vacancy mediated charge transport in α-$Cr_2O_3$ and Fe doped α-$Cr_2O_3$ systems. Both interstitial and vacancy-mediated diffusions are anisotropic with faster diffusion along (001) - direction. It is pointed out that hopping of charge carriers between two sites in α-$Cr_2O_3$ is energetically favourable when preferential orientation of the magnetic moment of the initial site (spin up state) is opposite to that of the final location (spin down state) due to AFM in-plane spin order of Cr ions. The migration of electrons from a "spin-up" site to a "spin-down" site, thus, involve a spin flip for Cr ions either in-plane or out of plane. This in contrast to charge hopping process in α-$Fe_2O_3$ structure, where an electron hopping is allowed between two Fe sites with parallel spin states and it is energetically favourable only for FM ordered in-plane Fe ions and non-favourable for AFM ordered inter-planes. Hence, magnetic spin order could play a significant role in the electrical properties of Cr doped α-$Fe_2O_3$ system. The DFT calculations showed a correlation of ferrimagnetic (FiM) ordering with reduction of band gap, Fe–O d–p orbital hybridization, and AFM type Fe–Cr σ type superexchange interaction in case of α-$Fe_{0.40}Cr_{1.60}O_{2.92}$ nano-structured system [29]. A first-principle study by Wang et al. [43] showed band gap reduction in the range of $(Fe_{1-x}Cr_x)_2O_3$ solid solutions, which cover our Cr doped system. The band gap reduction was attributed to the onset of Cr →Fe d−d excitations and broadening of the valence band due to hybridization of O 2p states with Fe and Cr 3d states. The interactions between Fe($3d^5$) and Cr ($3d^3$) ions take into account the relative concentration, spatial distribution of Fe and Cr ions and electronic structure in the solid solution. Especially, sitting of the Cr $t_{2g}$ states at the top of valence band (VBM) during initial increase of the Cr concentration makes the VBM energetically higher than that in pure α-$Fe_2O_3$ or α-$Cr_2O_3$. According to this calculation, modified magnetic spin order in the Cr doped α-$Fe_2O_3$ system depending on the spatial arrangement of Fe and Cr atoms can affect the electronic properties. We believe the decrease of inter-ionic distance by increasing the Cr content in α-$Fe_2O_3$ plays a major role for the enhancement of charge hopping process and enhancement of the d-shell electron spin coupling within the metal ions (Fe/Cr) pairs.

## 4. Conclusions

We studied structure, magnetic properties, and electrical conductivity in Cr doped α-$Fe_2O_3$ system. Magnetic properties in the as-alloyed samples are affected by magnetic heterogeneity at the interfaces of α-$Fe_2O_3$ and α-$Cr_2O_3$ phases. The soft ferromagnetic nature in as-alloyed samples is completely different from the parent α-$Fe_2O_3$ and α-$Cr_2O_3$ oxides. The as-alloyed samples indicated signature of two Morin transitions, which is suppressed in heat treated samples, especially at higher Cr content. It is attributed to modified magnetic spin structure and reduction of anisotropy in $Fe_{2-x}Cr_xO_3$ system. A local distortion in lattice structure and its coupling with spin order introduce first order magnetic phase for the samples with higher Cr content. Magnetic field sensitive spin-flipping process of Cr ions is prominent in the CAFM state of heat treat samples. The bulk magnetization in the material decreased on increasing the Cr content, where as the long range spin order at the Fe sites (form Mössbauer spectroscopy) is not much affected by Cr doping. The electrical conductivity in $Fe_{2-x}Cr_xO_3$ system is well explained by small polaron hopping of charge carriers. There is a possibility of fractional ($0 \leq \delta \leq 1$) charge transfer between trivalent Fe ($Fe^{(3-\delta)+} \leftrightarrow Fe^{(3+\delta)+}$) and Cr ($Cr^{(3+\delta)+} \leftrightarrow Cr^{(3-\delta)+}$) ions through $Fe^{3+}$-$O^{2-}$-$Fe^{3+}$, $Cr^{3+}$-$O^{2-}$-$Cr^{3+}$ and $Fe^{3+}$-$O^{2-}$-$Cr^{3+}$ superexchange paths. The enhancement of electrical conductivity on increasing Cr content in the material is associated with band gap reduction, where lattice defects, spatial arrangement of Fe and Cr ions, the hybridization of O 2p states with Fe and Cr 3d states, and coupling between Fe($3d^5$) and Cr ($3d^3$) ions can play a significant role.

**ACKNOWLEDGMENTS**

Authors acknowledge the measurement facilities in CIF, Pondicherry University for dc magnetization and Raman spectra, UGC-DAE CSR, Indore for Mössbauer spectra, and Indus-2, RRCAT, Indore (Synchrotron X-ray diffraction). RNB also acknowledges research support from UGC-DAE-CSR Indore centre (Grant No. CSR-IC-245/ 2017-18/1326).

**Figure captions**

Fig. 1 Refined SXRD patterns of as-alloyed samples of $Fe_{2-x}Cr_xO_3$ (x=0.2-0.8) compositions.

Fig.2 Refined SXRD patterns of the 1000 $^0$C heated samples of $Fe_{2-x}Cr_xO_3$ (x=0.2-0.8).

Fig.3 Raman spectra with peak profiles fitted using Lorentzian shape for the as-alloyed samples (a-c) and the samples heated at 1000 $^0$C (d-f).

Fig. 4 ZFC(T) and FC(T) curves at 100 Oe (a-e). Normalized ZFC(T) and FC(T) curves (f-g), difference (DM) between FC and ZFC curves (h), and their derivative curves (i).

Fig. 5 MZFC(T) and MFC(T) curves at 100 Oe (a-e) and normalized MZFC curves at different magnetic fields for two samples (f, g). The kink is marked by arrows.

Fig.6. M(H) curves of the as-alloyed samples at different temperatures (a-e). Insets clarify loops at 10 K and 300 K, and irrversibility effect at 10 K. The shift of FC loops with respect to ZFC loops (f-h).

Fig. 7 Comparative M(H) loops at 10 K (a), 300 K (b) and 200 K (c) for the as-alloyed (ASP) and heated at 800 $^0$C (A8h12), 1000 $^0$C (A10h12) samples. The ZFC and FC loops at 10 K are compared for A10h12 sample (d).

Fig. 8 Comparative plots of the temperature dependence of field ($H_C$ and $H_{exb}^{ZFC}$(a-c)) and magnetization ($M_R$ and $M_{exb}^{ZFC}$ (d-f)) parameters for the as-alloyed (ASP) and heated (at 800 $^0$C (A8h12), 1000 $^0$C (A10h12)) samples.

Fig. 9 The M(H) loop at 300 K for different samples (a-e). The initial M(H) curves at 10 K (f-j) and corresponding first order derivative (blue curves) for different samples. The $M^2$ vs. H/M curves based on initial M(H) curves of the samples at 10 K (f-j).

Fig. 10. Variation of $H_C$ and $M_R$ with Cr content in the as-alloyed (ASP) samples (a-b) and in the heat treated (A10) samples (c-d) at different measurement temperatures. The temperature dependence of $M_S$ (e) and $H_{sf}$ (f) for the A10 samples.

Fig. 11 Room temperature Mössbauer spectra for the as-allyoed samples (a-c) and the samples heated at 1000 $^0$C (d-h). The lines indicate components of fit data.

Fig.12 Electric field dependence of dc current density for different samples (a) and dc electrical conductivity calculated at selected electric fields (b).

**Table 1. The structural parameters from Rietveld refinement of the two-phased SXRD patterns of as- alloyed samples.**

| Sample | Atom | Position | | | Unit cell Parameters | | | Phase (%) |
|---|---|---|---|---|---|---|---|---|
| | | x | y | z | $a$ (Å) | $c$ (Å) | V(Å)$^3$ | |
| FeCr2MA (Mixed Phase) | Fe | 0 | 0 | 0.3539(5) | 5.0384 | 13.7686 | 302.70 | 88.87 |
| | O | 0.3169(6) | 0 | 0.25 | | | | |
| | Cr | 0 | 0 | 0.3492(3) | 4.9641 | 13.5970 | 290.17 | 11.13 |
| | O | 0.295(2) | 0 | 0.25 | | | | |
| FeCr4MA (Mixed Phase) | Fe | 0 | 0 | 0.3538(6) | 5.0351 | 13.7648 | 302.22 | 73.26 |
| | O | 0.3174(6) | 0 | 0.025 | | | | |
| | Cr | 0 | 0 | 0.3472(5) | 4.9609 | 13.6025 | 289.92 | 26.74 |
| | O | 0.3097(9) | 0 | 0.25 | | | | |
| FeCr5MA (Mixed Phase) | Fe | 0 | 0 | 0.3541(5) | 5.0168 | 13.7171 | 298.99 | 71.63 |
| | O | 0.3159(6) | 0 | 0.25 | | | | |
| | Cr | 0 | 0 | 0.3479(1) | 4.9396 | 13.5401 | 286.11 | 28.37 |
| | O | 0.3060(10) | 0 | 0.25 | | | | |
| FeCr6MA (Mixed Phase) | Fe | 0 | 0 | 0.3537(8) | 5.0171 | 13.7162 | 299.01 | 77.69 |
| | O | 0.3143(9) | 0 | 0.25 | | | | |
| | Cr | 0 | 0 | 0.3468(2) | 4.9456 | 13.5626 | 287.29 | 22.31 |
| | O | 0.3083(15) | 0 | 0.25 | | | | |
| FeCr8MA (Mixed Phase) | Fe | 0 | 0 | 0.3513(8) | 5.0149 | 13.7292 | 299.02 | 60.32 |
| | O | 0.3157(12) | 0 | 0.25 | | | | |
| | Cr | 0 | 0 | 0.3472(1) | 4.9455 | 13.5611 | 287.24 | 39.68 |
| | O | 0.3089(8) | 0 | 0.2500 | | | | |

**Table 2. The structural parameters from Rietveld refinement of the single phased SXRD patterns of the samples heated at 1000 ºC.**

| samples | Atom | x | y | z | $a$ (Å) | $c$ (Å) | c/a | V(Å)$^3$ | Grain size |
|---|---|---|---|---|---|---|---|---|---|
| FeCr2MA10 | Fe | 0 | 0 | 0.3545(3) | 5.0394 | 13.7407(4) | 2.727 | 302.28 | 35 nm |
| | O | 0.3081(6) | 0 | 0.2500 | | | | | |
| FeCr4MA10 | Fe | 0 | 0 | 0.3542(3) | 5.0284 | 13.6926(7) | 2.723 | 299.82 | 33 nm |
| | O | 0.3053(3) | 0 | 0.2500 | | | | | |
| FeCr5MA10 | Fe | 0 | 0 | 0.3535(7) | 5.02(6) | 13.6817 (2) | 2.725 | 299.06 | 30 nm |
| | O | 0.3024(6) | 0 | 0.2500 | | | | | |
| FeCr6MA10 | Fe | 0 | 0 | 0.3535(4) | 5.0121 | 13.6371 (2) | 2.721 | 296.72 | 34 nm |
| | O | 0.3060(4) | 0 | 0.2500 | | | | | |
| FeCr8MA10 | Fe | 0 | 0 | 0.3528(4) | 5.0047 | 13.6105 (5) | 2.719 | 295.23 | 34 nm |
| | O | 0.3054(4) | 0 | 0.2500 | | | | | |

**Table 3.** The band positions in Raman spectra for the samples heated at 1000 ºC of $Fe_{2-x}Cr_xO_3$ system.

| Cr (x) | $E_g(1)$ (cm$^{-1}$) | $E_g(3)$ (cm$^{-1}$) | $E_g(4)$ (cm$^{-1}$) | $A_{1g}(2)$ (cm$^{-1}$) | $E_g(5)$ (cm$^{-1}$) | $E_u$ (cm$^{-1}$) |
|---|---|---|---|---|---|---|
| 0.2 | 242.05±0.45 | 301.50±0.34 | 423.20±0.83 | 515.07±0.29 | 610.00±0.32 | 670.58±0.22 |
| 0.5 | 236.13±0.60 | 296.85±0.22 | 418.10±0.33 | 509.10±0.24 | 611.30±0.26 | 666.90±0.26 |
| 0.8 | 231.40±0.05 | 296.50±0.06 | 420.81±0.63 | 503.60±0.12 | 612.45±0.12 | 663.09±0.18 |

**Table 4.** The line parameter values from fit of the room temperature Mössbauer spectra for as-alloyed (ASP) and 1000 ºC heated (A10) samples of $Fe_{2-x}Cr_xO_3$ system.

| Sample name | Sextet component | FWHM (mm/s) | IS (mm/s) | QS (mm/s) | $B_{hf}$ (T) | Relative area (%) | $\chi^2$ |
|---|---|---|---|---|---|---|---|
| FeCr2MA | 1 sextet | 0.44±0.02 | 0.37±0.05 | -0.19±0.01 | 51.11±0.50 | 57.70±0.05 | 1.78 |
|  | 2 sextet | 0.42±0.08 | 0.37±0.09 | -0.18± 0.03 | 47.98±0.10 | 42.23±0.04 |  |
| FeCr4MA | 1 sextet | 0.47±0.02 | 0.37±0.01 | -0.18±0.02 | 51.04±0.06 | 50.61±0.05 | 1.35 |
|  | 2 sextet | 0.51±0.02 | 0.38 ±0.06 | -0.25± 0.03 | 46.64±0.12 | 49.39±0.01 |  |
| FeCr8MA | 1 sextet | 0.44±0.03 | 0.38 ±0.01 | -0.23±0.01 | 51.07±0.07 | 46.20±0.37 | 4.10 |
|  | 2 sextet | 0.49 ±0.06 | 0.37 ±0.01 | -0.24±0.01 | 46.70±0.02 | 53.01±0.31 |  |
| FeCr2MA10 | 1 sextet | 0.49±0.06 | 0.36±0.02 | -0.19±0.01 | 48.25±0.02 | 100±0.06 | 1.56 |
| FeCr4MA10 | 1 sextet | 0.45±0.02 | 0.37±0.02 | -0.24±0.01 | 50.54±0.07 | 82.72±0.08 | 0.98 |
|  | 2 sextet | 0.34±0.08 | 0.40±0.01 | -0.20±0.04 | 48.91±0.21 | 17.28±0.02 |  |
| FeCr5MA10 | 1 sextet | 0.44±0.02 | 0.36±0.01 | -0.22±0.01 | 50.76±0.08 | 53.96±0.09 | 1.05 |
|  | 2 sextet | 0.50±0.03 | 0.38±0.01 | -0.21±0.01 | 48.92±0.13 | 46.04±0.07 |  |
| FeCr6MA10 | 1 sextet | 0.49±0.02 | 0.37±0.01 | -0.22±0.01 | 50.00±0.03 | 59.50±0.23 | 2.40 |
|  | 2 sextet | 0.56±0.01 | 0.37±0.03 | -0.20±0.01 | 47.98±0.06 | 40.51±0.14 |  |
| FeCr8MA10 | 1 sextet | 0.39±0.01 | 0.36 ±0.01 | -0.23±0.01 | 51.45±0.01 | 64.40±0.20 | 1.43 |
|  | 2 sextet | 0.44±0.04 | 0.36±0.01 | - | 50.44±0.13 | 35.60±0.01 |  |

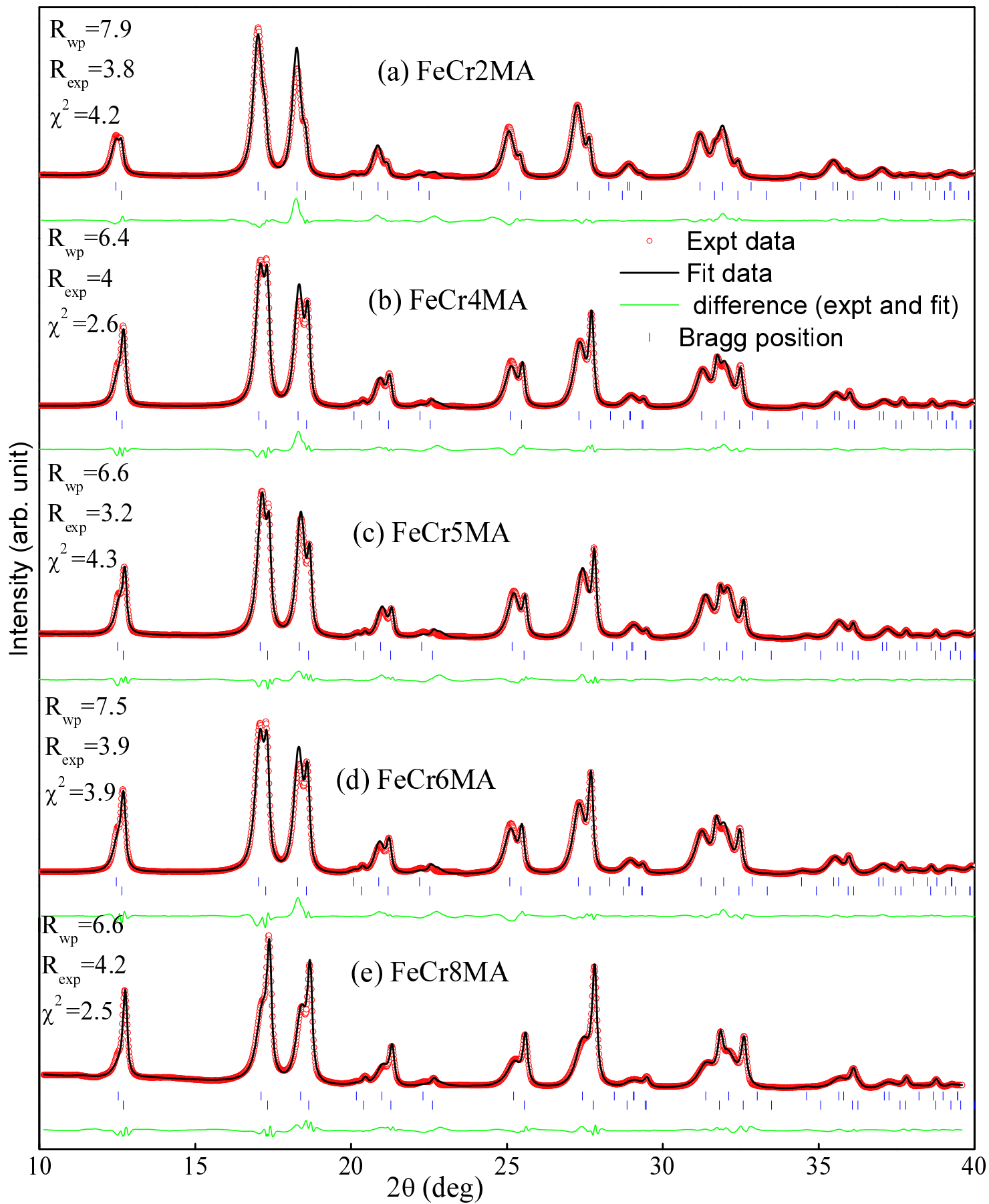

Fig. 1 Refined SXRD patterns of as-alloyed samples of $Fe_{2-x}Cr_xO_3$ (x=0.2-0.8) compositions.

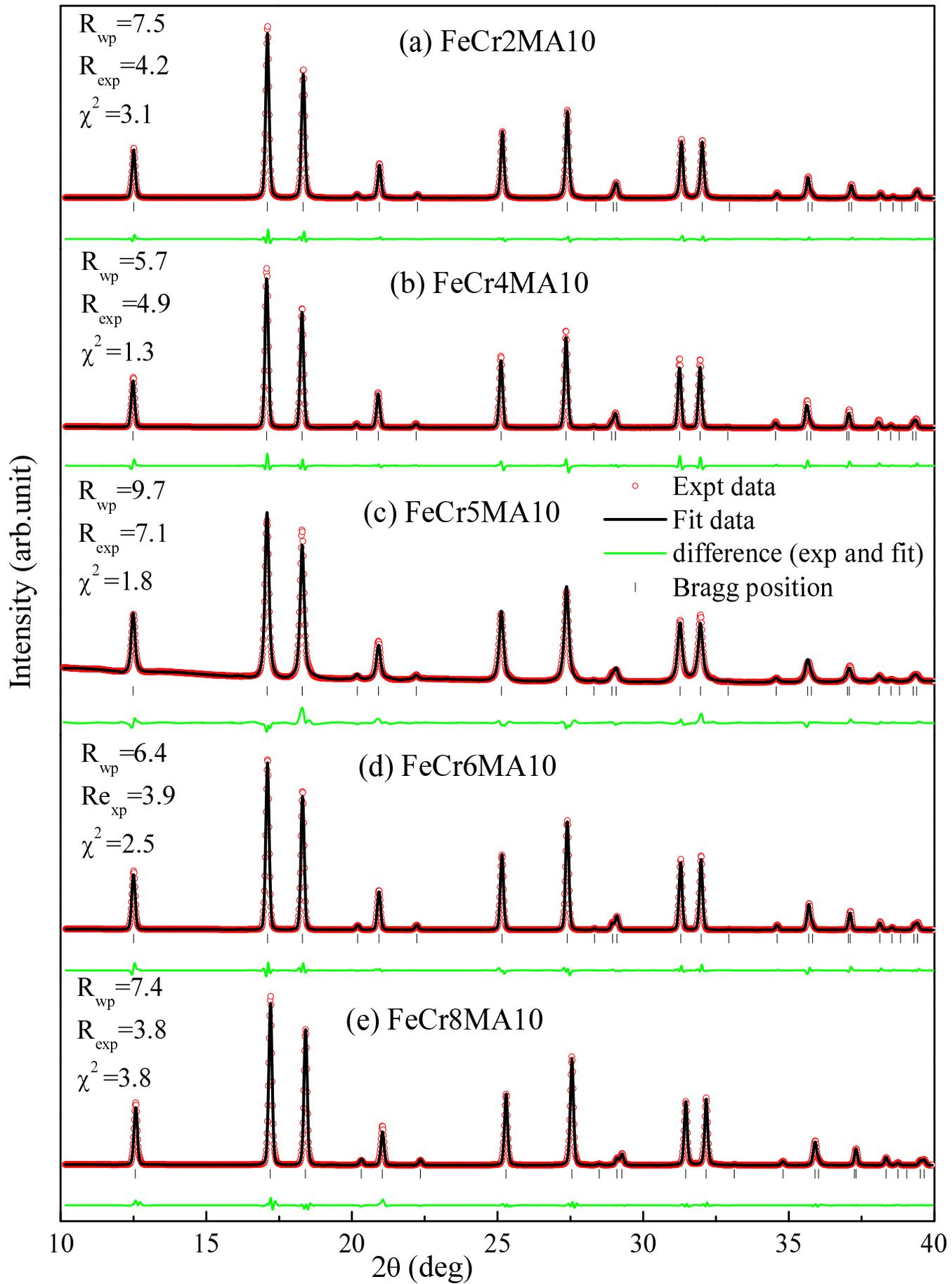

Fig.2 Refined SXRD patterns of the 1000 $^0$C heated samples of $Fe_{2-x}Cr_xO_3$ (x=0.2-0.8).

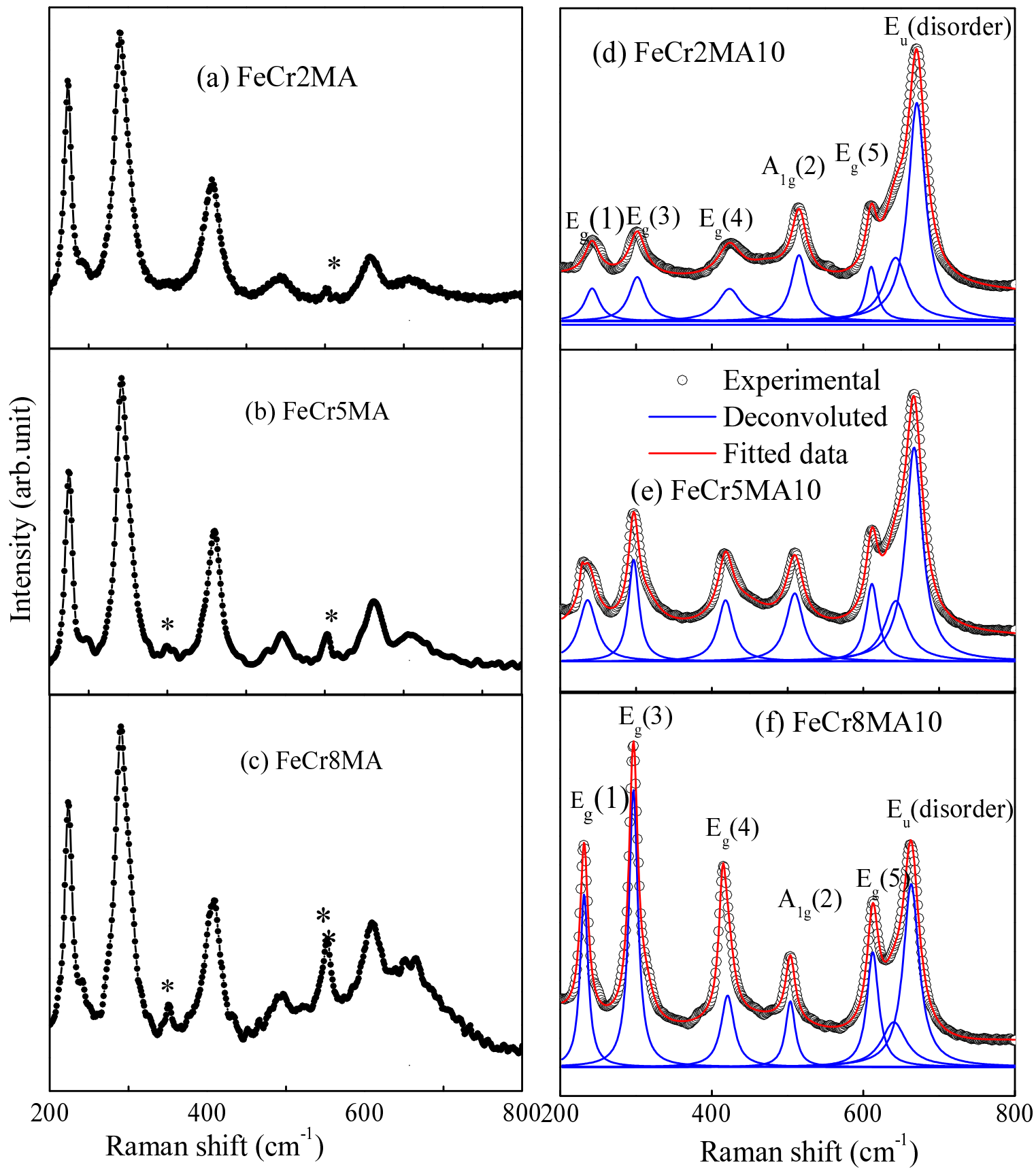

Fig.3 Raman spectra with peak profiles fitted using Lorentzian shape for the as-alloyed samples (a-c) and the samples heated at 1000 $^0$C (d-f).

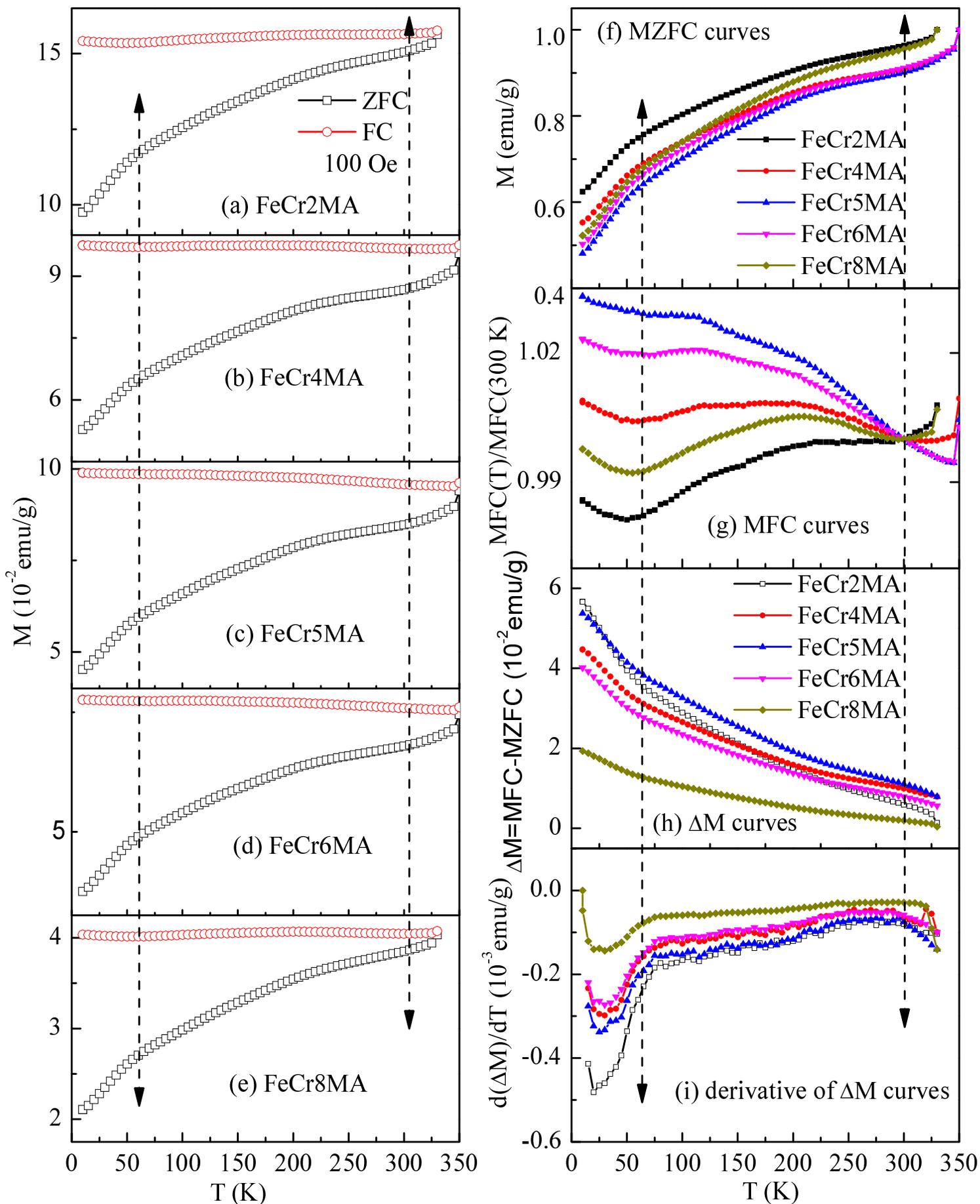

Fig. 4 ZFC(T) and FC(T) curves at 100 Oe (a-e). Normalized ZFC(T) and FC(T) curves (f-g), difference (ΔM) between FC and ZFC curves (h), and their derivative curves (i).

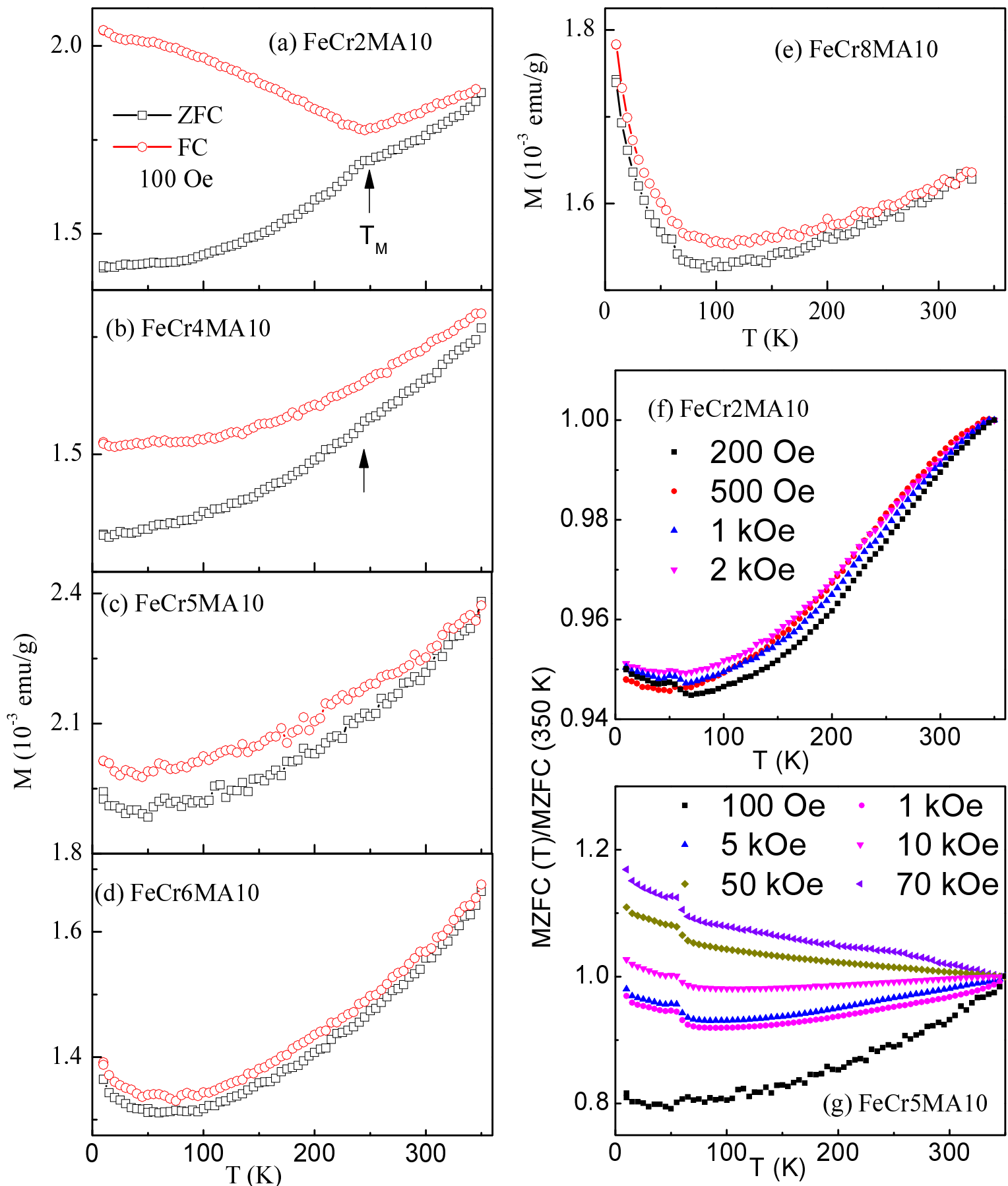

Fig. 5 MZFC(T) and MFC(T) curves at 100 Oe (a-e) and normalized MZFC curves at different magnetic fields for two samples (f, g). The kink is marked by arrows.

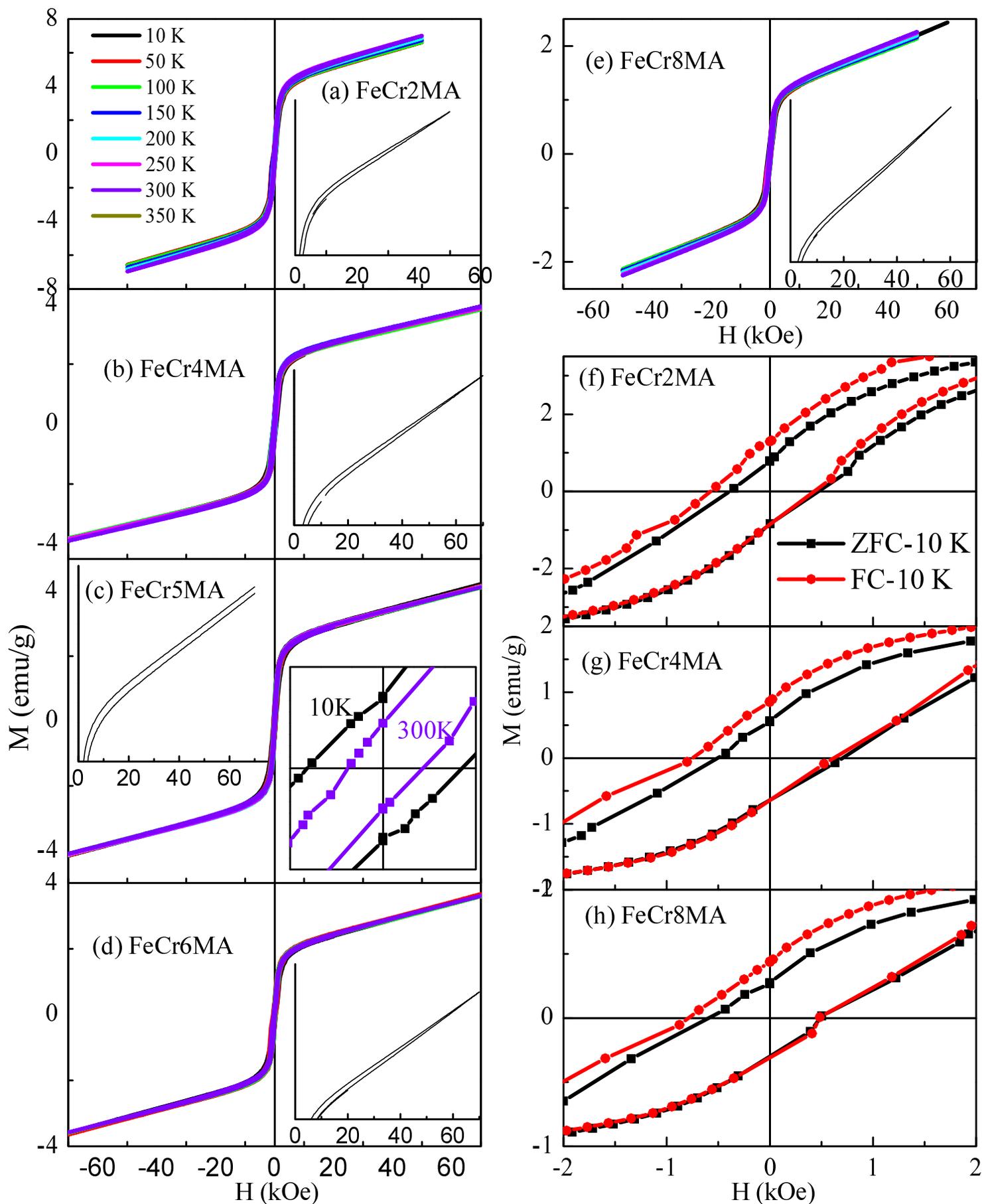

Fig.6. M(H) curves of the as-alloyed samples at different temperatures (a-e). Insets clarify loops at 10 K and 300 K, and irreversibility effect at 10 K. The shift of FC loops with respect to ZFC loops (f-h).

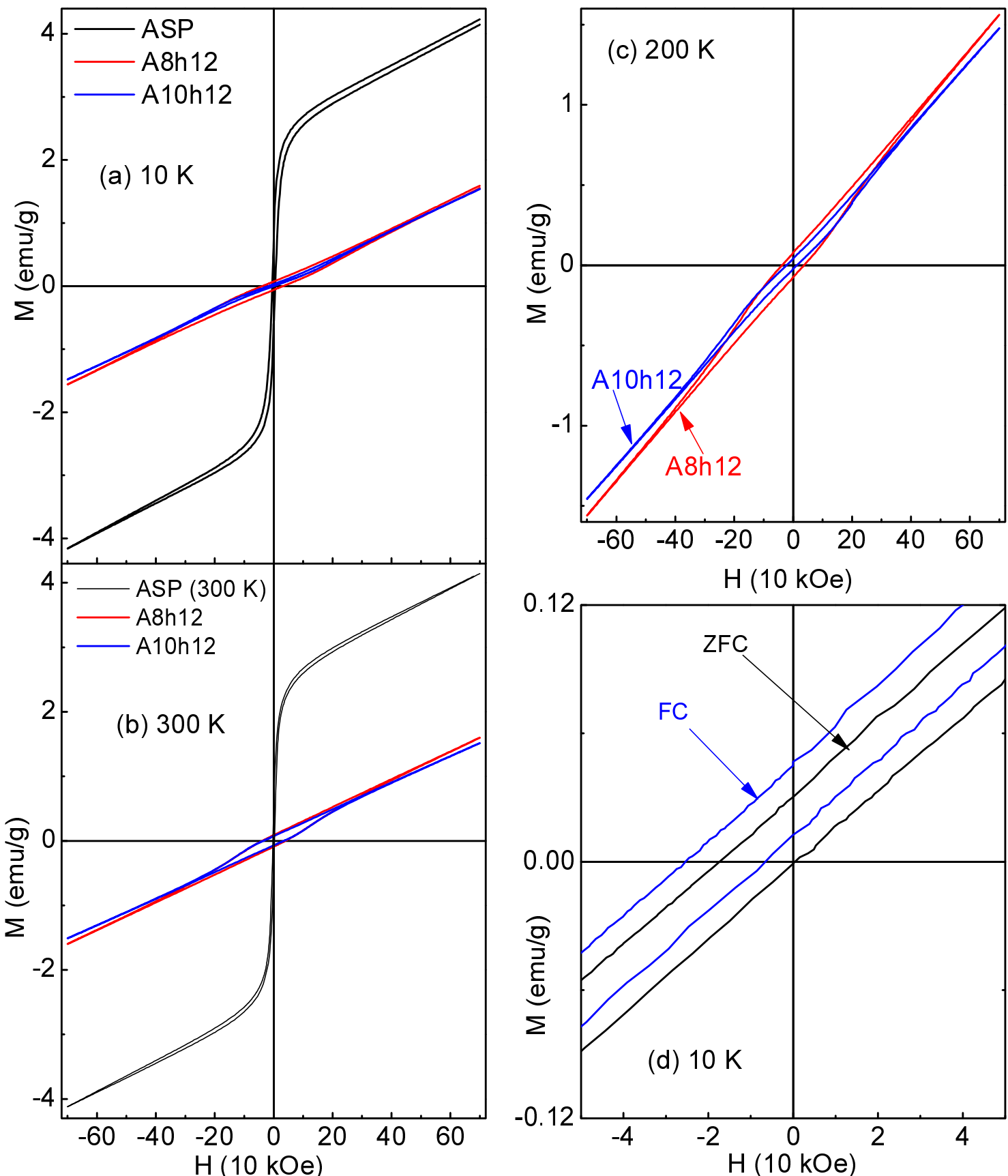

Fig. 7 Comparative M(H) loops at 10 K (a), 300 K (b) and 200 K (c) for the as-alloyed (ASP) and heated at 800 °C (A8h12), 1000 °C (A10h12) samples. The ZFC and FC loops at 10 K are compared for A10h12 sample (d).

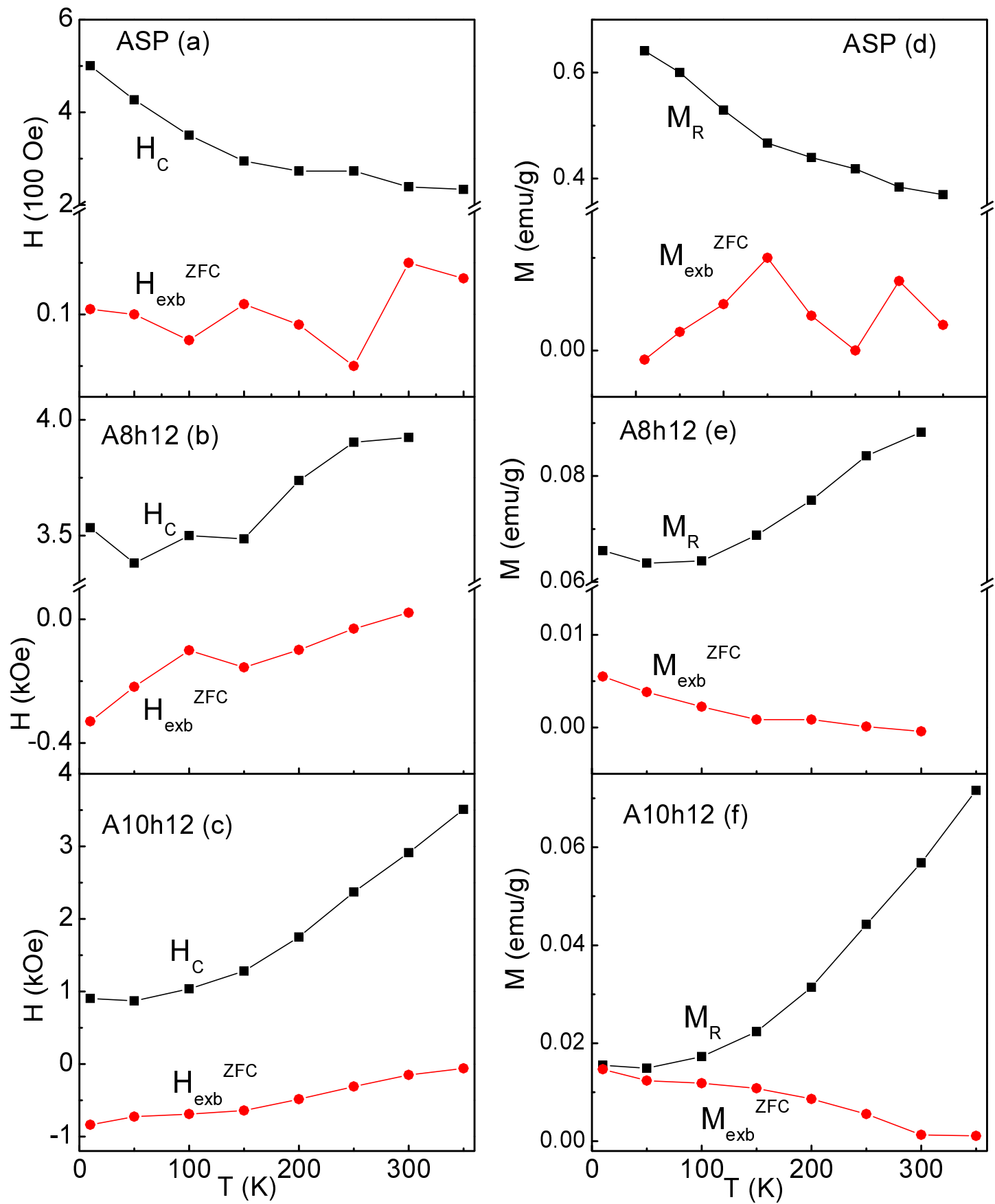

Fig. 8 Comparative plots of the temperature dependence of field ($H_C$ and $H_{exb}^{ZFC}$ (a-c)) and magnetization ($M_R$ and $M_{exb}^{ZFC}$ (d-f)) parameters for the as-alloyed (ASP) and heated (at 800 $^0$C (A8h12), 1000 $^0$C (A10h12)) samples.

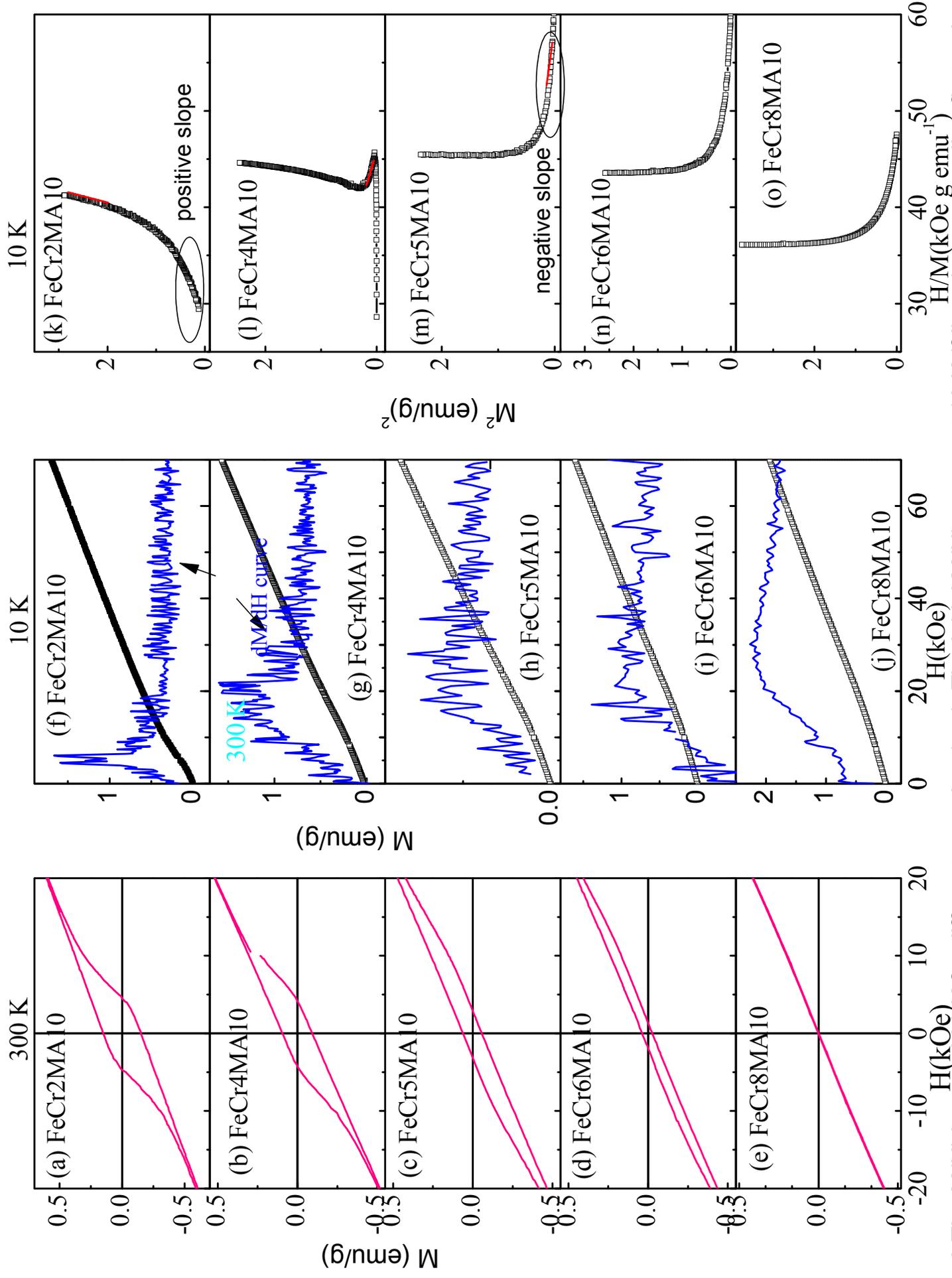

Fig. 9 The M(H) loop at 300 K for different samples (a-e). The initial M(H) curves at 10 K (f-j) and corresponding first order derivative (blue curves) for different samples. The $M^2$ vs. $H/M$ curves based on initial M(H) curves of the samples at 10 K (f-j).

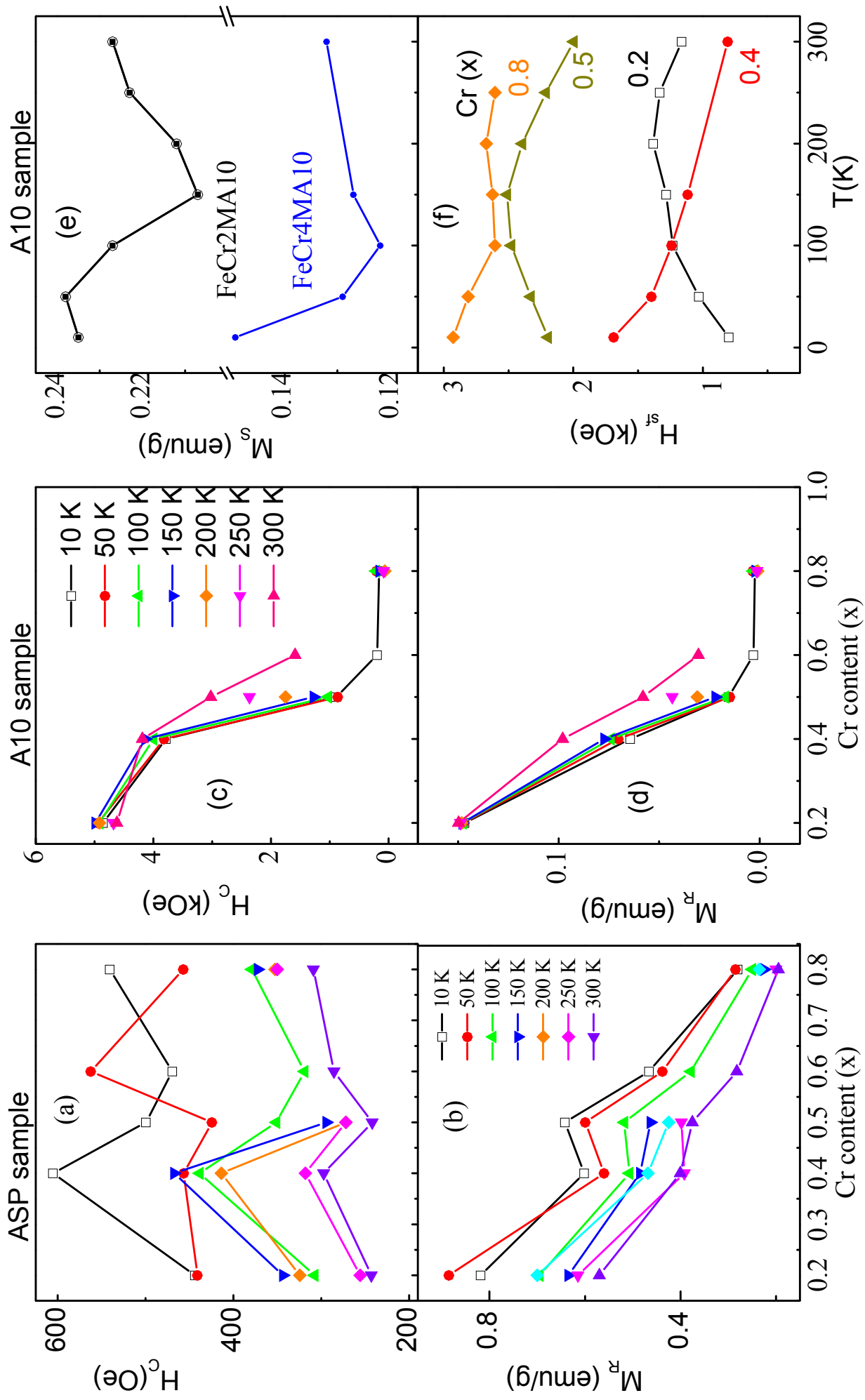

Fig. 10. Variation of $H_C$ and $M_R$ with Cr content in the as-alloyed (ASP) samples (a-b) and in the heat treated (A10) samples (c-d) at different measurement temperatures. The temperature dependence of $M_S$ (e) and $H_{sf}$ (f) for the A10 samples.

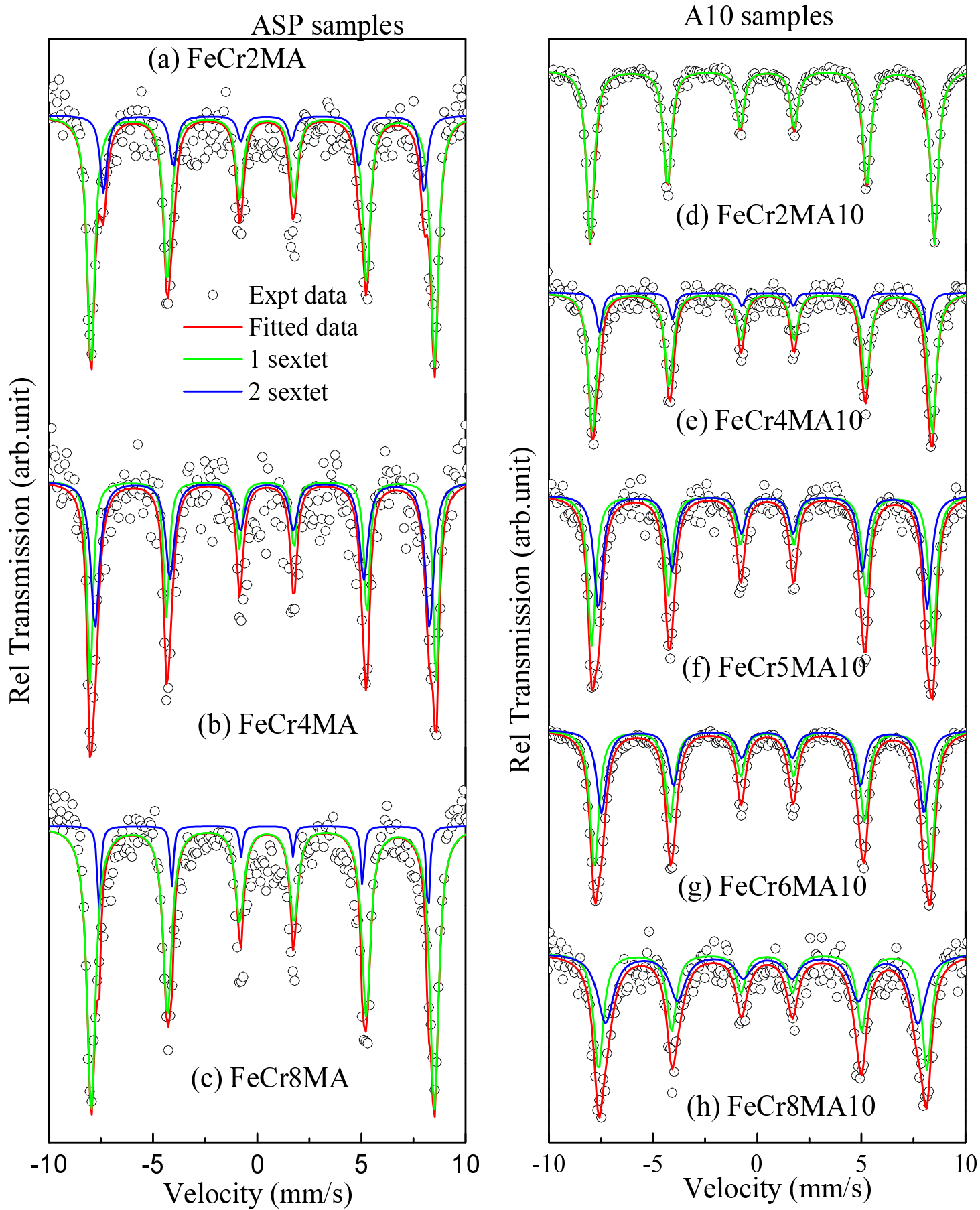

Fig. 11 Room temperature Mössbauer spectra for the as-allyoed samples (a-c) and the samples heated at 1000 $^0$C (d-h). The lines indicate components of fit data.

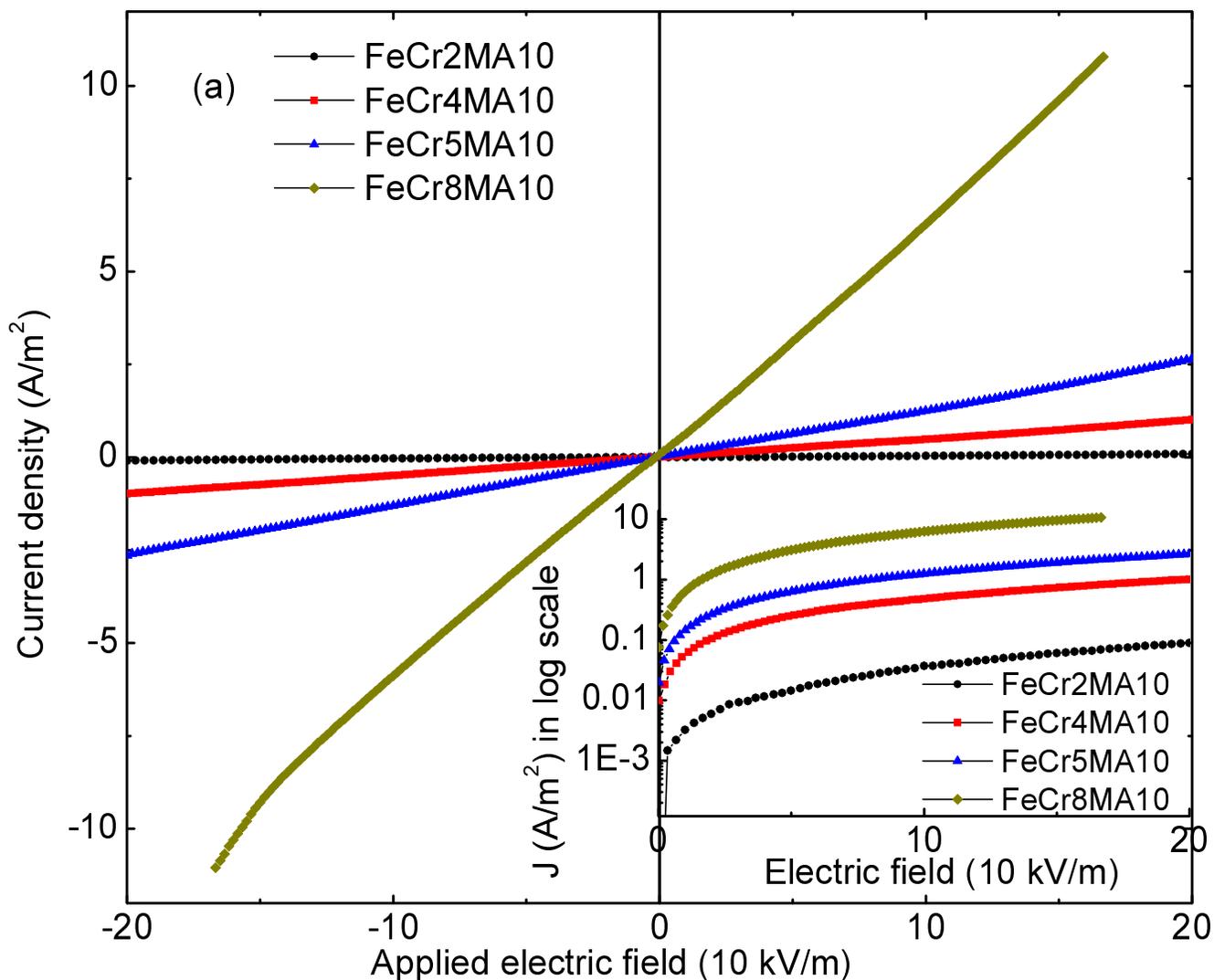

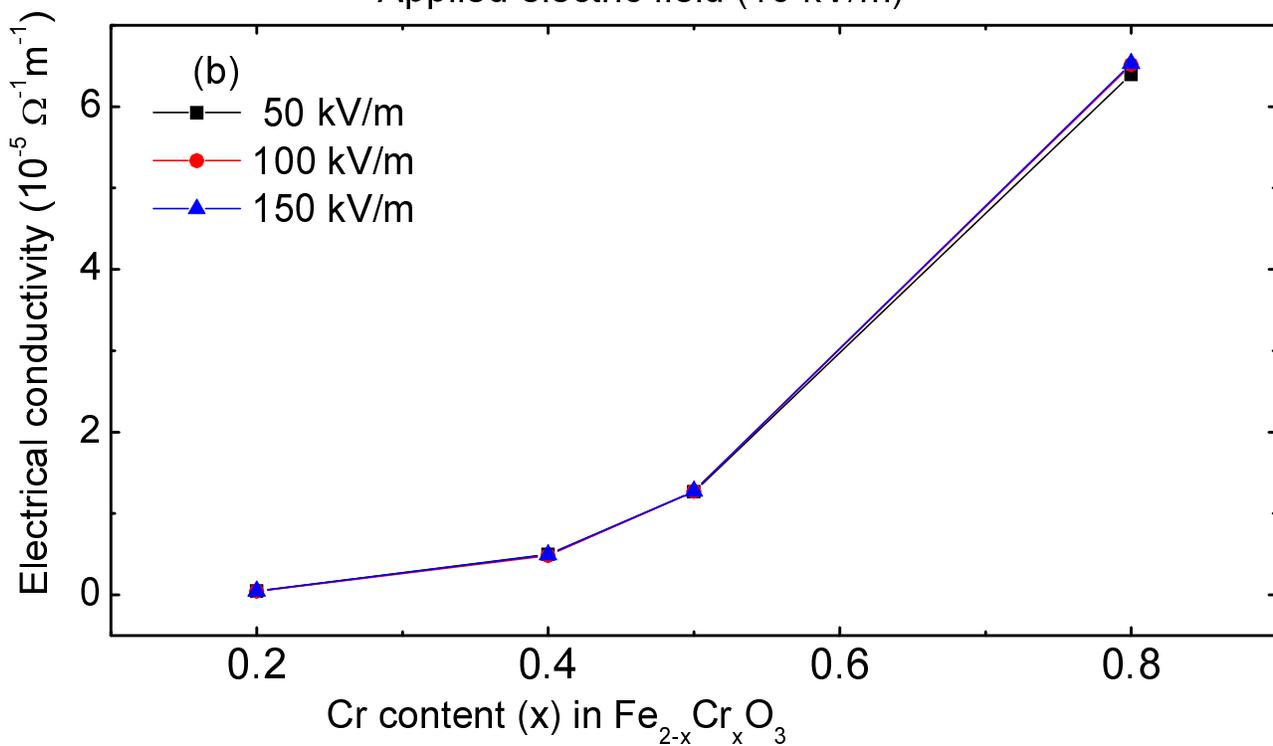

Fig.12 Electric field dependence of dc current density for different samples (a) and dc electrical conductivity calculated at selected electric fields (b).